# Chronological Analysis of Rigvedic Mandalas using Social Networks

Shreekanth M Prabhu
*Computer Science and Engineering Department,*
*Cambridge Institute of Technology*
Bengaluru,Inida
0000-0001-8154-9625
Email:shreekanthpm@gmail.com

Gopalpillai Radhakrishnan
*Computer Science and Engineering Department)*
*CMR Institute of Technology*
Bengaluru, India
radhakrishnan.g@cmrit.ac.in

*Abstract—* **Establishing the chronology of the Vedas has interested scholars for the last two centuries. The oldest among them is Rig-Veda which has ten Mandalas, each composed separately. In this paper, we look at deciphering plausible pointers to the internal chronology of the Mandalas, by focusing on Gods and Goddesses worshiped in different Mandalas. We apply text analysis to the Mandalas using Clustering Techniques based on Cosine Similarity. Then we represent the association of deities with Mandalas using a grid-based Social Network that is amenable to chronological analysis and demonstrates the benefits of using Social Network Analysis for the problem at hand. Further, we analyze references to rivers to arrive at additional correlations. The approach used can be deployed generically to analyze other kinds of references and mentions and arrive at more substantive inferences.**

I. INTRODUCTION

The Rig Veda is divided into ten volumes (Mandalas). These Mandalas, in turn, are made up of distinct dynasties of seers that have existed for ages, if not millennia. The Rig Veda is followed by Sama Veda, Yajur Veda, and Atharva Veda. The Vedas are followed by the Brahmanas, Aranyakas, and Upanishads, in which the emphasis shifts from rites to philosophical inquiry. The Vedas share many similarities with the Avesta in Iranian tradition.

Max Muller [1] popularized Vedic studies in Europe. Max Muller postulated a date of 1500 BCE for the Vedas, which became a 'historical fact" simply by repetition by other scholars. Closely tied to this was the position that composers of Veda were not autochthonous to India and rather they came via the Northwest (Russian Steppes is one origin) into India. Both of these were at odds with how Indians viewed their history, where they believed they were indeed native to India with no memory of migration into India from outside and vivid memories of all internal migrations within India.

In this context, the internal chronology of Rig-Veda, the relative ordering of the Mandalas becomes all the more important to draw inferences, rather than viewing the Vedas as a monolith. Using linguistic research, German Vedic scholar Oldenburg [2] established an approximate relative order of chronology for the Mandalas (Volumes) of the Rig Veda. The viewpoints of various scholars from multiple disciplines regarding the chronology of the Rig Veda and the ancestry of the Vedic people have been gathered by Edward Bryant and Laurie Patton [3]. Currently, there is no single hypothesis on Vedic chronology is accepted by all scholars as historians, language experts, archaeologists, geneticists, and Sanskrit scholars disagree with each other.

As far as Indian tradition goes, it is the Puranas and not the Vedas that are historical accounts. F.E. Pargiter [4] studied the inter-relationship between Vedas and Purana and tried to draw synchronisms. In doing so, he deviated from the general tendency of Western scholars to validate Indian chronology primarily using external markers. According to Pargiter, the Vedic people are a tribe that originated in India's Gangetic plain and then migrated westward. He also finds parallels in Purana to one of India's tribes (Druhyus) migrating towards the Northwest (to Europe?), carrying similar traditions with them.

The Vedic Gods seem to have crossed over from India to Europe, encompassing Celtic, Germanic, Nordic, Baltic, Slavic, Roman, Greek, Iranian, and a few more traditions. Dumezil [5] has made a detailed study on this. Everywhere, including India, more Gods and Goddesses have been incorporated into the pantheon over time, some unique to certain localities only. The final country to convert to Christianity in the 14th century was Lithuania, which has possibly kept more of the customs. We have accounts of the Zend and Avesta in the case of Iran, which shares many characteristics with the Vedas.

MacDonnell [6] conducted a thorough analysis of Vedic Mythology a century ago. Roshen Dalal [7] provides a concise account of the Vedic Gods. Numerous epithets, patronyms, and metronyms are used to refer to Vedic deities. The same God can also go by many names, and the same name can be used for different Gods in various situations. Additionally, gods are worshiped in pairs, triads, and groups. There is mutuality between Gods. Different Mandalas occasionally present varying accounts of the Gods. Poets and seers extensively combine truth and metaphor in this intricate, extensive body of literature. Ram Swarup [8] has done seminal work on the names of Vedic Gods and their attributes.

The Vedic Gods are also mentioned in Jain and Buddhist Literature. Even in contemporary Hindu Society, every Kula(clan) has its deity, referred to as Kula Devata. This is in addition to many other Gods and Goddesses they commonly worship. In a country with substantial internal migration, it is the Gods and Goddesses they worship that give identity to a community, more than language, locality, profession, or anything else. Thus, Gods unite Indians across linguistic groups, The same phenomenon is observed if we look at Egyptian, Greek, and European pantheons, where there is a commonality that cuts across all other differences.

Rig Veda contains ten Mandalas. Each Mandala is a book of hymns. Each hymn (Sukta) comprises several verses (Riks). Usually, a hymn is dedicated to a single God, duals, and/or groups of Gods. Generally, the object of prayer in a given hymn is referred to as 'Devata'.

In this paper, the scope of our analysis is the relative internal chronology of Rig-Vedic Mandalas. Thus, we compare references across Mandalas. The objective of the paper is to develop methodologies for textual analysis of Rig-Vedic Mandalas. We focus on Devatas in different Mandalas to do this analysis. Certain Gods are worshiped as duals and we factor these separately. The data source for the counters is work by Roshen Dalal [7]. We do similarity analysis followed by agglomerative clustering, where term frequencies of Devatas are used to compare Mandalas. We propose a novel grid-based social network representation to elicit insights for the chronological analysis of Mandalas. This paper succeeds our earlier work [9], where affiliative social networks were used.

The remaining portions of this paper are structured as follows. The Literature Survey is covered in Section II. Section III gives an overview of Vedic Devatas. details the methodology we used for textual analysis. Section IV covers the Analysis of Rigvedic Mandalas. Section V, Conclusions, concludes the paper.

## II. Literature Survey

### A. External Chronology of Rigveda

Possehl and Witzel [10] consider Vedas to be after the mature Harappan period, around 1000-1200 BCE. They arrive at these dates by tracing the transformation of Sanskrit from a codified form of Panini to an archaic form found in Vedas. Another marker they use is the introduction of Iron. Rigveda does not refer to Iron. It refers to Ayas which these scholars consider as Copper. A few decades back the known date for the Introduction of Iron was 1200 BCE or so. They however accept that the Vedas were largely composed in Punjab, where even Sindh was relatively distant.

Then there are Hittite and Mitanni references to Gods such as Ashvins which is dated between 1400 to 1600 BCE. Witzel considers Hittite languages more archaic than Vedic and then claims Vedic has to be later.

Such late dates for Rigveda which have been made starting with Max Muller are coming into question. Firstly, Panini's date itself is uncertain. Further, Sanskrit is a very different language compared to other natural languages. The decadence theories that apply to colloquial languages which change rapidly due to lack of grammatical foundation and strong knowledge-based institutional structure to preserve languages will not apply to Sanskrit. This point of view is argued well by Dattaraj Deshpande [11]. Sanskrit has been a formal language with thousands of scholars all over India collectively preserving it for centuries and millennia. Sanskrit has the unique ability to retain multiple forms of words for the same concept/object/entity. The spread of Sanskrit words has influenced not only Indo-European Languages but also Dravidian Languages. Swaminatha Aiyar [12] argues that many day-to-day Dravidian Language words have roots in Vedic Sanskrit.

K Suresh [13] who made a detailed study of Vedas details the gigantic scholarly tradition of Vedas in India. He divides the creation of Vedic Literature into 4 periods: (i) The period of creation of mantras, (ii) The period of collection of Mantras (iii) the Brahmana period, and (iv) the Sutra period. The seers involved were numerous spanning multiple generations generating mammoth literature. The scholarly genealogies span 50-100 generations. All this adds credence to the inference that the Vedic literature would have taken millennia to evolve and Vedic Sanskrit indeed is ancient. Over and above this Vedas are interpreted using a variety of perspectives. Ram Gopal [14] published a significant work that describes the inherent complexities involved. Even common words such as 'go(cow)' and 'ashva(horse)' may mean very differently depending on the context as Vedas are replete with allusions and metaphors.

The archaeologists [3] have not detected any discontinuity in the population in geographies mentioned in Rigveda to account for any influx, rather there is continuity going back many millennia. According to geologists, the Sarasvati River dried up by 1900 BCE. As the Vedas describe Sarsvati as a full-flowing river going into the sea and that phenomenon is more likely around 3000 BCE, the dates given by Western Scholars for Vedas do not seem credible. There is also a common inherited memory over generations of Saraswat Brahmins across India who claim they used to stay on the branch of Sarasvati and then dispersed to other parts of India.

Semenenko [15] correlated archeological findings i.e. material culture in various Harappan sites with textual references in Rigveda and arrived at a timeline for Rigveda from 3300 BCE to 2600 BCE. The latest findings in Sinauli have discovered a chariot, weapons, and a whip used for horses. This is dated 2200-1900 BCE. There are news reports [16] about archeological findings that attest to the use of Iron in Tamil Nadu in 2172 BCE.

Amitabha Ghosh [17] analyzed the astronomical observations referred to in Vedic texts such as stellar conjunctions, eclipses, equinoxes, solstices as well as exaltation of planets such as Mars. These observations were picked from Vedas, Brahmanas, and associated literature and the plausible dates are arrived at using modern astronomical software. He lists 5 observations for the pre-Vedic and early Vedic periods: 8326 BCE, 4677 BCE, 4539 BCE, 4350 BCE, and 4105 BCE. For the Vedic period, he lists 6 observations: 3961 BCE, 3928 BCE, 3541 BCE, 3281 BCE, 2948 BCE, and 2924 BCE.

Dr Niraj Rai [18], a geneticist rules out immigration into India. Rather provides plausible evidence for steady emigration out of India. He has publications in the pipeline to established journals regarding this.

Discounting all these and just going by the judgment of Western linguists and compressing the timelines to just a few centuries may be not wise.

*B. Internal Chronology of Rigveda*

Scholars agree that Rigveda was followed by Sama Veda then Yajurveda and Atharva Veda was the last, allowing for some degree of chronological and textual overlap.

According to Elst [19] "A collection of hymns related to the psychedelic brew Soma forms book 9, and a distinctly younger collection of hymns constitutes book 10. This latter is part of a younger culture shared with the Yajur- and Atharva-Veda (the Samaveda mostly consists of hymns of the Ṛigveda put to music). The Yajur-Veda reaches down to the age of the dynasty's fraternal war, related in the Mahābhārata, ("great [epic] of the Bhārata clan"), and the youngest layer of the Rigveda likewise, mentioning king Śantanu, the great-grandfather of the war's protagonists, in hymn 10:98. It was their grandfather (or Śantanu's stepson) Kṛṣṇa Dvaipayana, a.k.a. Veda Vyasa who closed the Vedic corpus by giving it its definitive structure. The last king mentioned in the Vedic corpus is Vyasa's biological son Dhṛtarāṣṭra, father of the Kaurava participants in the battle".

All scholars agree that Mandala 10 is late and family Mandalas 2-7 are earlier Mandalas. The relative chronology of Mandalas 1,8 and 9 is rather uncertain and has been often disputed. Ditrich [20] covers different viewpoints on the chronology of scholars such as Wust, Bergaigne, Lanman, Hopkins, Arnold, and Oldenberg,

Talageri [21] who did a textual analysis of Rigveda hypothesized that the order of Mandalas is 6,3,7 which he calls Early Mandalas, followed by 4,2 and later books 5,8,1,10. Table 1, details the Internal Chronology of Rigveda as per Talageri. Here the analysis is based on the genealogy of Kings and Rishis, further corroborated by geographical analysis, among many other things.

**Table 1: Internal Chronology of Rigveda as per Talageri**

| Mandala 6 Mandala 3 Mandala 7 Mandala 1 | Mandala 4 Mandala 2 Mandala 5 | Mandala 8 | Mandala 9 | |
|---|---|---|---|---|
| | | | | Mandala 10 |

The import of Talageri's work is that the earliest Mandalas in his ordering have references to the eastern region and later Mandalas have references to Northwestern India. This is contrary to the Aryan Influx theory, where Aryans were supposed to have come from the Northwest and then steadily moved towards the Ganges plains over time.

*C. Research Opportunity*

The traditional Indian scholars are firm that both Vedas and the Vedic people are native to India. They do not concern themselves with examining the commonality between Indian and European traditions. At the same time western scholars in particular linguists, are emphatic that the Aryan Language and people originated outside India, as part of a larger Indo-European family. Closely tied with this hypothesis is the chronology of Vedic Mandalas. The date for Rigveda given by Western linguists around 1500 BCE has become mainstream even when it is fiercely contested by other scholars.

Chronological problem is studied by scholars from multiple disciplines. There is an opportunity for scholars from Computer Science to dwell on the problem, as part of an emerging field of "Digital Humanities". In this paper, we analyze text using clustering algorithms and then represent the data using social networks. The visual power of social networks can represent chronological insights far more powerfully.

### III. VEDIC DEVATAS

Numerous Gods, Goddesses, and stories are shared by Celtic, Germanic, Nordic, Baltic, Slavic, Roman, Greek, Iranian, and Vedic traditions. In some cases, even the titles are the same or similar, whereas in others, the concepts are similar. Lithuania, the final country to be converted to Christianity in the 14th century, is likely to have maintained more of the shared tradition. In the case of Iran, we have Zend and Avesta narratives, which have similarities to the Vedas in numerous respects. Table 2 below gives a list of Vedic Gods and a brief description. Table 3 gives known correspondences among traditions.

**Table 2: Vedic Gods**

| Name | Nature of God |
|---|---|
| Agni | Associated with Fire, sacrifice, messenger to the Gods |
| Indra | Thunder God and King of Gods |
| Aditi | Mother of Gods. Liberates from bond |
| Varuna | God of rivers. Preserves cosmic order |
| Mitra | God of light, Friendship |
| Aryaman | Invoked during oath for marriage |
| Pushan | Protector on routes, forests, pastoral God |
| Ashvins | Twin Gods, Associated with sea and horse |
| Parjanya | Another name for Indra, Thunder |
| Ushes | Goddess of dawn |
| Dyaus | Sky-god. Generally paired with Prithvi. Earth-God |
| Bhaga | Aditya like Varuna, Benefactor |
| Soma | Plant/drink/God. |
| Vayu | Wind-God |
| Rudra | Rigvedic Deity corresponds to the later deity Shiva. Rudra means roarer. One who eradicates evil from roots. |
| Vishnu | Intervenes as a close friend of Indra. In Puranas, one who maintains the universe Pervasive God |
| Savitra | impeller, rouser, vivifier, associated with Sun God |
| Surya | Sun God |
| Prajapati | Creator God |
| Yama | God/King associated with death |
| Purusha | Cosmic being from whom the world gets generated |
| Danu | Her progeny is named as Danavas. Also known as Diti whose progeny are Daityas |
| Ida/ILa | Goddess, Mother of many tribes |
| Manu | First Man who is also a King |

**Table 3 Vedic Gods in other Indo-European traditions**

| Vedic God | Other Indo-European Traditions |
|---|---|
| Dyaus(Pitar) | Zeus(Greek), Jupiter(Latin), DeiPatrous(Illyrian) |

| Usha(Ushes) | Eos(Greek).Aurora(Roman), Ausra(Baltic), Ostara(Germanic), Ausrine(Lithuanian). |
|---|---|
| Parjanya (Indra) | Perkunas(Lithuanian), Perkuns(Prussian), Perkons(Lativian),Perun(Slavic), Fjorgyn(Norse),Thor(Germanic), Sucellus(Celtic), Taranis (Celtic) |
| Bhaga | Bogu(Slavic) |
| Ashvins | Asvieniai(Lihuvenian), Romulus(Roman) Castor and Pollux(Greek), Hengst and Horsa(Germanic), Volos and Veles (Slavic) also worshipped by Celts |
| Varuna | Ahura Mazda, similar names in other IE traditions |
| Aryaman | Ariomanus(Gaulish), Airyaman(Iranian), Eiremon(Irish) |
| Mitra | Mithra(Iranian) |
| Pushan | Pan(Greek), Faurus(Roman) |
| Danu | Danu(Irish) |

Another prominent concept is the Group of Gods, the members may be siblings or they share the same property. For example, Adityas are Sun-Gods group among celestial Gods, associated with heaven and light. They have commonality yet their individuality. Savitr is more of an abstract concept whereas Surya is a physical manifestation. Please note that an epithet is a name derived from a property or attribute. Maruts are storm Gods. Rhibus are semi-divine beings who became Gods because of their skill. Table 4, details Groups of Gods.

**Table 4: Groups of Gods**

| Group Name | Members |
|---|---|
| Adityas | Varuna, Mitra, Aryaman, Amsha, Bhaga, ushan, Daksha, Martanda, Dhatr, Savitr, Tvastr, Vishnu, Vivasvat, Shakra |
| Maruts | Generally referred to as Group, without giving member names |
| Rhibus | Rhibhukshan, Vraja and Vibhuvan |
| Rudras | Ahi Bhutanya, Bhaga, Nirritti, etc. |
| Visva Devas | All Gods |
| Devas (33) | Vasus (8), Adityas (12), Rudras (11), Ashvins (2) |

.
Many Vedic Gods are worshipped as duals. Some of the prominent pairs are Mitra-Varuna and Dyaus-Prithvi. Indra, Varuna, Soma, Agni, Pushan, and the Maruts also get worshipped in duals.

There are three classes of Gods – Celestial, Aerial, and Terrestrial. But this can be a manifestation of the same phenomenon. Agni is a terrestrial God, associated with lightning in the air and the Sun in the sky, all are associated with fire and light. Originally "Asura" was the term used for great God and later developed negative connotations as competitors, opponents, and enemies of Devas. However, Devas, Asuras, Daityas, and Danavas form the same common ancestry.

## IV. ANALYSIS OF RIGVEDIC MANDALAS

### A. Clustering based on Similarity Analysis

We make use of a data set [22] that has counts of hymns dedicated to specific Devatas in specific Mandalas. Using the counts of Vedic Devatas across Mandalas, we arrive at Cosine Similarity. This data in turn is used to cluster Mandalas agglomeratively. Figure 1, depicts the clustering of all Mandalas on Gods worshipped (singly).

According to Talageri, the order of Vedic Mandalas is as follows. Early Mandalas are 6, 3,7. The middle Mandalas are 4,2. The late books are 5,8,1,10. Here Mandala 9 is exclusively dedicated to God Soma. The clustering analysis below can be used to draw inferences based on the commonality of Devatas worshiped

In Figure 1, the Mandalas 3 and 7 appear in nearby clusters. Mandala 6 may be a forerunner for Mandala 4. Mandalas 4 and 2 appear in clusters twice removed from each other. Mandalas 1 and 5 are in the same cluster. The late Mandala 8 and Mandala 10 are twice removed from each other, here. Thus, with single Gods' clustering analysis, there is partial correspondence as far as the chronology of Mandalas is concerned, but there is no clear distinction between early, middle, and late Mandalas

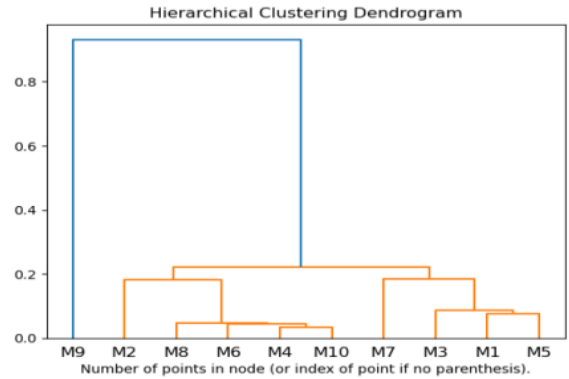

Figure 1: Cluster of Mandalas based on Single Gods/Devatas

Since it is well-accepted that the family Mandalas are earlier than non-family Mandalas. We apply the clustering algorithm now only to family Mandalas, to get greater clarity on interrelationships.

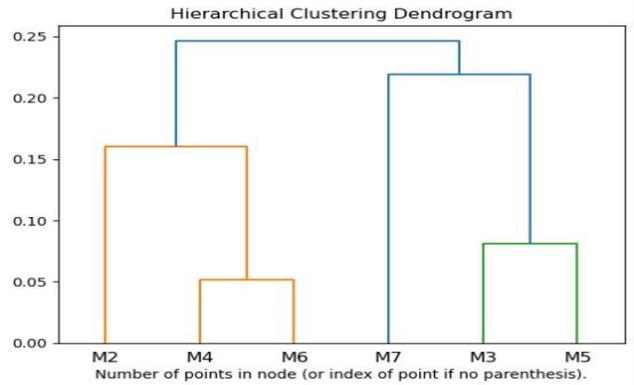

Figure 2: Cluster of Family Mandalas based on Single Gods/Devatas

The above visualization reconfirms the affinity of Mandalas 6, 4, and 2. Here Mandala 6 is early whereas 4 and 2 are middle Mandalas. It also confirms the affinity between Mandalas 3, 5, and 7. Mandala 3 can be considered a forerunner for Mandala 5. Thus, inferences from Figure 1 and Figure 2 are largely the same.

Next, we look at hymns dedicated to Duals in family Mandalas. Figure 3 illustrates the agglomerative clustering of Mandals based on such hymns. Here early Mandalas 6 and 7 are directly linked whereas Mandala 3 is once removed among early Mandalas. The middle Mandalas 4 and 2 are directly linked and once removed from early Mandala 3. Mandala 5, a later Mandala is directly linked to Mandala 6 and Mandala 7. These findings reflect a good degree of continuity across early and middle Mandalas as far as the deities worshiped are concerned, at a macro level.

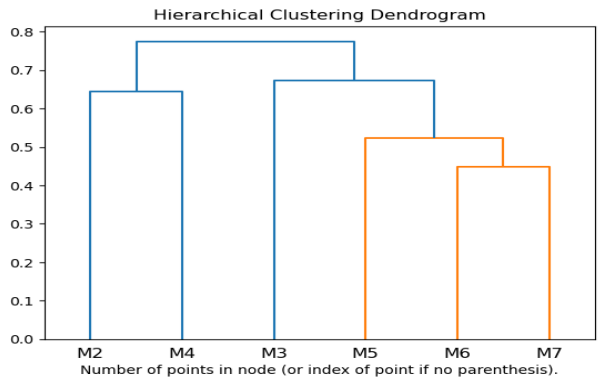

Figure 3: Cluster of Family Mandalas based on Duals

The data set [23] is based on Suktas and Riks that refer to the rivers in the Rigvedic Mandalas, here Sukta consists of varying numbers of Riks. Using the counts of rivers across Vedic Mandalas, we arrive at Cosine Similarity. This data in turn is used to cluster Mandalas agglomeratively. Figure 4, depicts the agglomerative clustering of Mandalas based on river mentions.

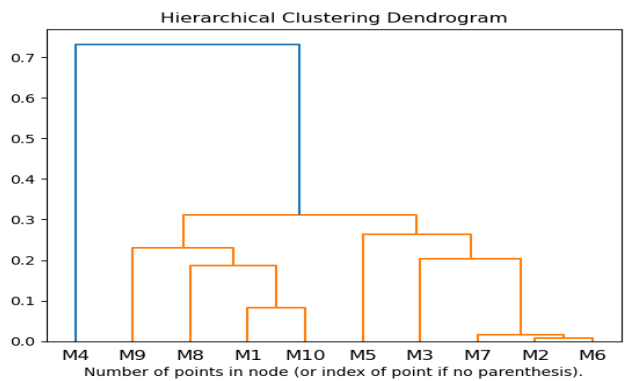

Figure 4: Cluster of Vedic Mandalas based on river mentions.

Here the geographic locales related to early Mandalas 6,3,7 are spread between three neighbouring clusters and then they are linked to Mandala 5. Mandala 4 appears in a stand-alone cluster. Mandala 1 and 10, which are voluminous and developed over long time horizons are in the same cluster. Mandala 8 which is an intermediate Mandala is linked to these late Mandalas, which in turn is linked to Mandala 9. The chronological sequence is more apparent here if we accept that the Vedic people moved from place to place or Vedic culture by involving more Rishis spread to newer geographies.

The same data set was subjected to K-Means analysis and generated 3 clusters: Cluster 1(Mandala 6,3,7 and 2), Cluster 2(1,5,9 and 10), and Cluster 3(Mandala 4 and 8). Cluster 1 has all early Mandalas and one Middle Mandala. Cluster 2 has Mandalas which have evolved over a long time, covering large regions. Cluster 3 has one middle Mandala and one non-family Mandala. The pattern of clustering reflects some geographic transition from east to northwest.

Whereas clustering techniques based on similarity analysis can help draw some inferences and validate a set of hypotheses, when it comes to visualization, they do not provide a convenient and elegant mechanism. With that in mind, we turn to Social Network Analysis.

### B. Social Network Analysis

Conventional social networks such as affiliation networks will be rather limited in extracting insights. We made use of such networks in an earlier work [9]. In this paper, we propose a novel grid network to trace how Suktas addressed to Devatas vary from Mandala to Mandala. Here Devatas as nodes are repeated vertically on the grid and Mandalas are depicted on horizontal axes of the grid.

Figure 5, represents Devatas mentioned in Mandalas. In this representation, we ignore the frequency of mentions. What we are representing is Devatas worshipped in Mandalas, as a vertical trace. Mandala 1 evolved over a long-time horizon concurrently with Middle Mandalas and later. We have not included a depiction that covers Mandala 10, as it has way too many unique Devatas not found in earlier Mandalas and accepted by all scholars as the youngest Mandala. These depictions indicate certain Gods as early Gods. In general, there is a continuity of Gods, which is eternal. No manifestation gets dropped.

Next, we look at duals. Figure 6, represents Duals in all Mandalas. Mandala 10 has no duals. Thus, duals as a concept lost traction in later Mandalas. Certain Gods such as Indra have duals with many other Gods exhibiting mutuality between Gods. Then certain duals such as Dyava-Prithvi found together only with each other. Mitra-Varuna is another dual, here Mitra is only found with Varuna, but not vice-versa.

From this analysis, we can infer that the Devatas are known from the earliest Mandalas and persist in the middle Mandalas as well as non-family Mandalas. This is because the core tenets that go into defining Devatas have conceptual coherence and they only manifest as concrete forms, abstract forms, and other forms. Further, there is a continuity of Gods in post-Vedic tradition. Ram Swarup [8] explains the names of Gods and the intent behind them.

It is also important to dispel false narratives and half-truths in textbooks and mainstream media. Generally, it is claimed that Vedic Gods Indra, Varuna, and Agni gave way to Vishnu, Shiva, and Brahma. In closer examination, it can be seen that Vishnu is referred to in the earliest Mandalas. Shiva is the beneficent form of Rudra, who is worshipped in Mandalas prominently. Prajapati in Vedas leads to Brahma later. More importantly, Gods such as Varuna, Indra, and Agni continue to be prayed in contemporary rituals. Thus, we are looking at a continuing tradition.

Another insight we get is the Vedic Gods as a collective are rather ancient and they moved to other Indo-European traditions long back and over long periods. At times same or cognate names were retained. At other times the names were completely different, Further, using similar processes newer Gods and Goddesses are visualized, manifested, and worshipped.

Next, we represent the geospatial grid network, here rivers appear on vertical lines and these vertical lines are arranged from the west to east. Thus, an eastern association of early Mandalas and a western(northwestern) association of later Mandalas becomes very apparent using this visualization. To make the analysis easy, here also we ignore the frequency of mentions. The progression from east to west is visualized well in Figure 7.

## V. CONCLUSIONS

Chronology of Vedas is a vexing question that has occupied the minds of scholars for centuries. External chronology has markers such as the date of drying of the Sarasvati River at one end and the earliest authenticated astronomical observation at the other. As and when new archeological findings surface, the earliest dates of Iron (not referred to in the Rigveda) and the horse and chariots (frequently mentioned in Rigveda) keep getting revised, in turn affecting the timelines of Rigveda. In this paper, we have focused on the internal chronology of the Rigvedic Mandalas. We attempt to corroborate the relative order of Mandalas given by Talageri [20] using hymns (Suktas) addressed to Vedic Devatas and rivers mentioned in Mandalas. We make use of agglomerative clustering followed by novel grid-based Social Network representation to exploit the power of visualization. We observe that there is a great degree of continuity of Devatas starting from the very first Mandala to younger Mandalas. In the youngest Mandala, many more Devata are worshipped, but earlier Devatas persist. Our findings corroborate with the work of Ditrich [20] who studied references to Dvandva(dual) compounds which included non-theonyms and dual theonyms and concluded that there is no marked difference in the way of usage across Mandalas to account for relative chronology with various stages. However, we conclude that there indeed is a relative chronology of Mandalas, but the theonyms reflect a largely continuing tradition.

When it comes to rivers, the Mandalas reflect an order from the east towards the west, with Sarasvati being the constant. All these allude to the possibility that Vedic tradition was a continuing tradition in India with internal migrations. Further, plausibly Vedic tradition has been mirrored in other traditions in a varied manner. The technique we have used can also be applied to references of animals, metals, and other objects as well. More broadly we have demonstrated the power of Social Networks for text analysis.

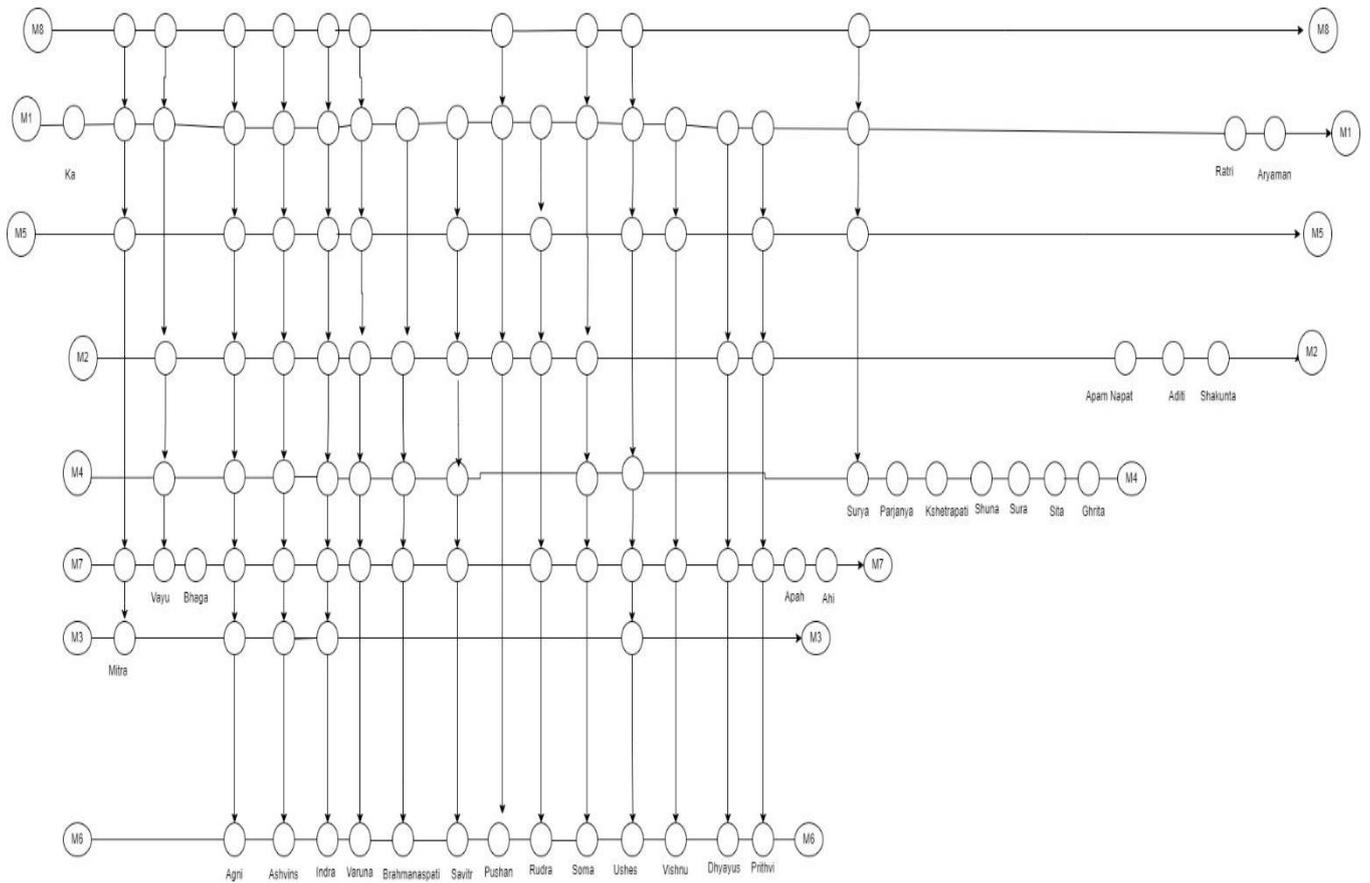

**Figure 5: Devatas in Mandalas**

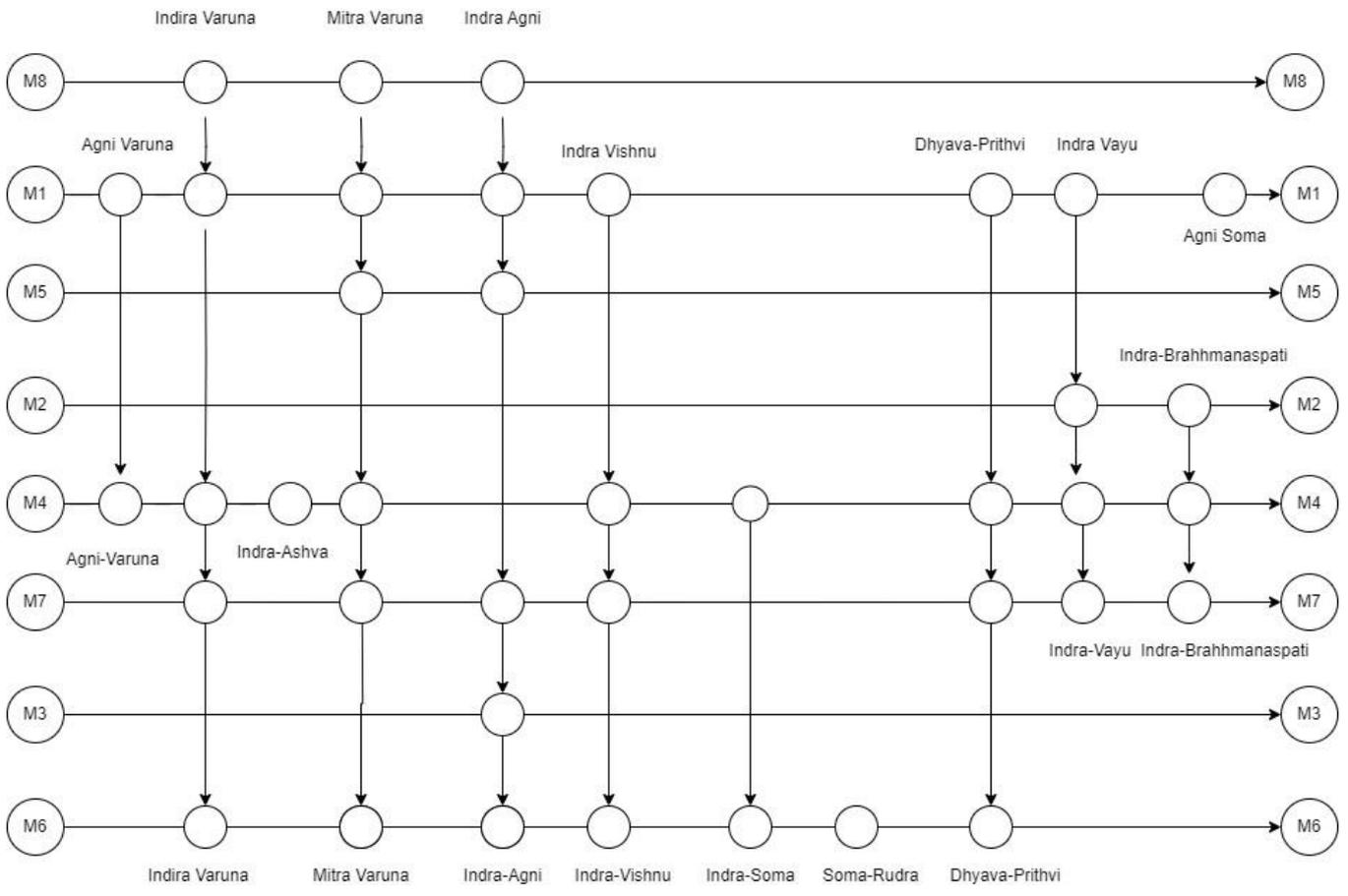

**Figure 6: Duals in Mandalas**

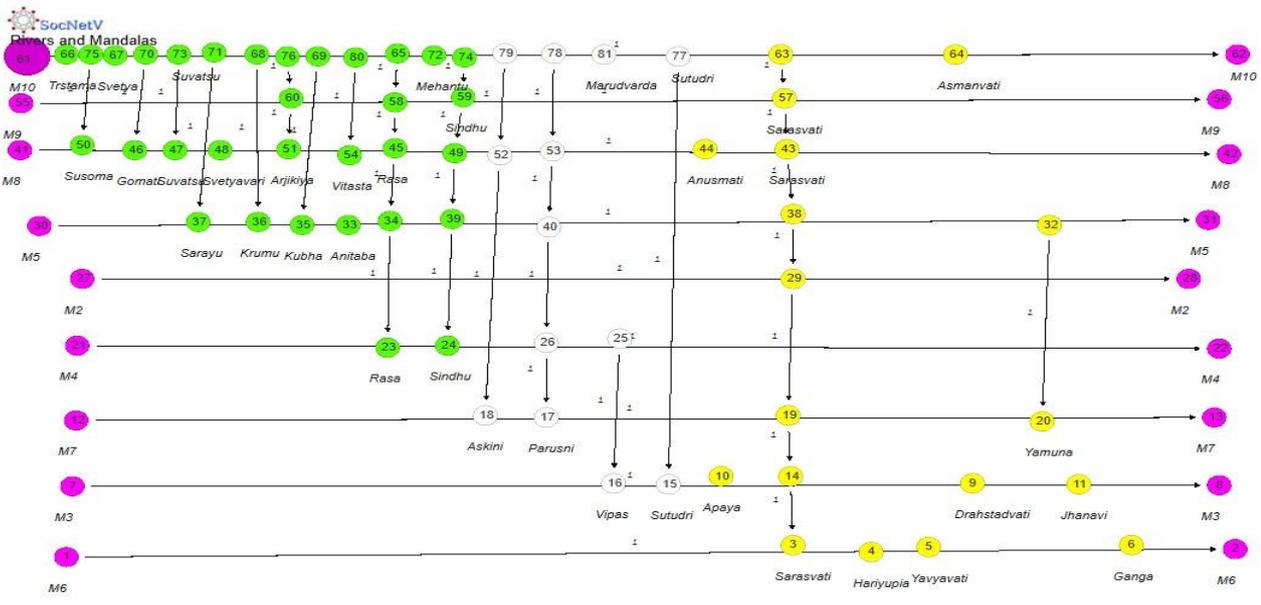

**Figure 7 Rivers in Vedic Mandalas**